\documentclass[%
 reprint, prx,
superscriptaddress,
 amsmath,amssymb,
 aps,
]{revtex4-2}
\usepackage{graphicx}
\usepackage{dcolumn}
\usepackage{bm}
\usepackage{textcomp} 
\usepackage{gensymb} 
\usepackage{xcolor}
\usepackage{ulem}
\usepackage{float}
\usepackage{physics}
\usepackage{siunitx}
\DeclareSIUnit{\Er}{E_{\text{rec}}}
\usepackage{hyperref}
\usepackage{tcolorbox}
\DeclareUnicodeCharacter{2212}{-}

\newcommand{\Vtp}{V_{\mathrm{TP}}}
\newcommand{\unkx}{u_{\alpha,\mathbf{k}}(\mathbf{x})}
\newcommand{\eikx}{e^{i\mathbf{k}\mathbf{x}}}
\newcommand{\kp}{\mathbf{k}_\mathrm{p}}
\newcommand{\kc}{\mathbf{k}_\mathrm{c}}
\newcommand{\kpl}{\mathbf{k}_+}
\newcommand{\kmi}{\mathbf{k}_-}
\newcommand{\Erec}{E_\mathrm{rec}}

\newcommand{\sico}{\sin(\mathbf{k}_\mathrm{p}\mathbf{x})\cos(\mathbf{k}_\mathrm{c}\mathbf{x})}

\newcommand{\coco}{\cos(\mathbf{k}_\mathrm{p}\mathbf{x})\cos(\mathbf{k}_\mathrm{c}\mathbf{x})}
\newcommand{\dc}{\Delta_\mathrm{c}}
\begin{document}

\preprint{APS/123-QED}

\title{Bandstructure of a coupled BEC-cavity system: effects of dissipation and geometry}

\author{David Baur}
\author{Simon Hertlein}
\author{Alexander Baumgärtner}
\author{Justyna Stefaniak}
\author{Tilman Esslinger}
\author{Gabriele Natale}
\email{gnatale@phys.ethz.ch}
\author{Tobias Donner}
\affiliation{Institute for Quantum Electronics, Eidgen\"ossische Technische Hochschule Z\"urich, Otto-Stern-Weg 1, CH-8093 Zurich, Switzerland}

\date{\today}

\begin{abstract}
We present a theoretical model for a transversally driven Bose-Einstein condensate coupled to an optical cavity. We focus on the interplay between different coherent couplings, which can trigger a structural phase transition, known as the superradiant phase transition. Our approach, based on band structure theory and a mean-field description, enables a comprehensive analysis of the nature of the system's excited modes, precursing the phase transitions. By incorporating dissipative couplings, intrinsic to these systems, we find non-Hermitian phenomena such as the coalescence of crossing precursor modes and the emergence of exceptional points (EPs). The general formulation of our model allows us to explain the role of an angle between transverse pump and the cavity deviating from $90^\circ$. This offers us a unified perspective on the plethora of different implementations of such systems.
\end{abstract}

\maketitle

\section{Introduction}
Many-body cavity quantum electrodynamics (QED) provides a powerful platform for studying collective quantum phenomena, where light mediates interactions between particles~\cite{Mivehvar2021Cavity,Defenu2023Long}. A key realization of such a system is a Bose-Einstein condensate (BEC) coupled to an optical cavity and illuminated transversally by a laser field. The physics explored in such setups lies at the intersection of quantum optics and condensed matter, giving rise to a rich array of observable phenomena, including structural phase transitions to super- and subradiant phases~\cite{baumann2010,Kollar2017Supermode,Wolf2018Observation, Baumgartner2024Stability, helson2023density, zhang2021observation}, the formation of supersolids~\cite{leonard2017supersolid,Schuster2020Supersolid,Guo2021Optical}, dynamic spin-orbit coupling~\cite{kroeze2019dynamical}, or the realization of lattice models with competing short- and long-range interactions~\cite{landig2016,Klinder2015Observation}. More recently, the interplay and competition between coherent and dissipative couplings in these open many-body systems has been explored, leading to observations of dissipation-stabilized phases~\cite{Ferri2021Emerging} and dissipation-induced instabilities~\cite{dogra2019,buvca2019dissipation,dreon2022,booker2020non,Kongkhambut2024Observation,Kongkhambut2022Observation}. These phenomena have been studied across a variety of implementations, including different geometries, atomic species, cavity types, and cavity dissipation rates~\cite{Mivehvar2021Cavity}. 

A common hallmark of the studied phase transitions is the softening of collective excitation modes as the system approaches a phase boundary. These modes serve as precursors to the self-ordered phases that emerge at the transition: even in the unordered phase, the system exhibits density fluctuations that mirror the spatial structure of the ordered phase~\cite{Schneider1972Phase,Santos2003Roton}. At the critical point, the excitation energy of these modes vanishes, leading to strong fluctuations that ultimately break a symmetry~\cite{Sachdev2000Quantum}. Examining these excitation modes in both the unordered and ordered phases provides key insights into the nature of the phase transition and the underlying mechanisms driving it~\cite{mottl2012,landig2015,leonard2017monitoring}.

\begin{figure}[h!]
\centering
\includegraphics[width=0.48\textwidth]{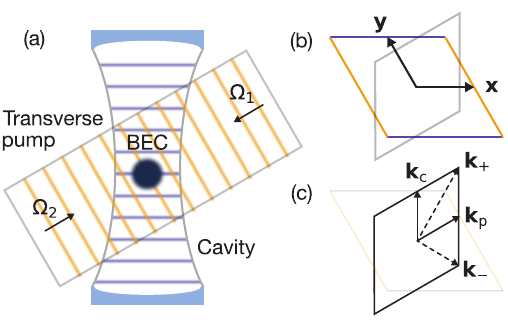}
\caption{\textbf{BEC-Cavity system with transverse pump.} (a) Schematic illustration of a Bose-Einstein Condensate (BEC) coupled to a single-mode optical cavity. The BEC is transversally illuminated by two counter propagating light fields with Rabi rates $\Omega_1$ and $\Omega_2$. These fields form a standing wave lattice with a residual running wave component, determined by the imbalance parameter $\gamma=\sqrt{\Omega_1/\Omega_2}$. For this illustration, the angle between the transverse beams and the cavity mode is $\theta = 60^\circ$. Panel (b) shows the real-space unit cell (Wigner-Seitz cell), which is chosen to contain one period of the transverse pump and cavity lattice. Due to the interference of the transverse pump and cavity fields, an additional lattice emerges with half the periodicity. Panel (c) shows the corresponding reciprocal unit cell (Brillouin zone), which is spanned by the two $\lambda/2$ periodic lattices of the system, 
the cavity wavevector $\kc$ and transverse pump wavevector $\kp$, where $\lambda=2\pi/\mathbf{k}_\mathrm{p,c}$. We highlight the two vectors $\kpl$ and $\kmi$, defined as $\mathbf{k}_\pm:=\kp\pm\kc$.}
\label{fig:setup}
\end{figure}

In systems with competing orders -- where multiple spatial density structures can form but are in competition -- the simple picture of a single softening excitation mode no longer holds~\cite{Morales2018Coupling}, requiring a more involved description~\cite{Demler2004Theory}. In the companion paper~\cite{Natale2026Synchronization}, we experimentally investigate such a scenario, spectroscopically resolving the interplay between two distinct excitation modes. Our findings reveal that in the presence of dissipation, a non-Hermitian coupling between these modes can lead to their coalescence and synchronization.
While numerical solutions of the Gross–Pitaevskii equation (GPE) can provide detailed simulations of such systems~\cite{GPE_SciPostPhysCodeb}, they often lack a conceptual notion of the mechanisms at play. In this work, we develop a theoretical framework based on band structure theory and mean-field approximations to offer both qualitative insights and a quantitative description of the excitation modes and their synchronization. 
Although our results are derived for a specific setting, our model extends to a variety of configurations, including different geometries (e.g., a varying angle between the transverse pump and cavity, and an imbalance between counterpropagating transverse pumps), cavity dissipation, or sign of the optical potentials~\cite{Roux2020Strongly,Kessler2014Steering,Kollár2015Adjustable,leonard2017supersolid,zhang2021observation,baumann2010}. Notably, we provide an intuitive explanation for the effects of a transverse pump-cavity angle different from 90$^\circ$, and establish a foundation for further exploration of non-Hermitian phenomena in these systems.

This article is structured as follows: In Section~\ref{sec:system}, we introduce the system under study -- a BEC coupled to an optical cavity and transversally driven by a dispersively detuned light field. We then present a brief introduction to band structure formalism and adapt it to our setting, providing the foundation for analysing the density-modulated BEC in the optical lattices formed by the various light fields. Building on this, we develop a mean-field (MF) theory that offers an intuitive description of the system and enables us to recover the phase diagram. Our analysis reveals two excitation modes that serve as precursors to the two possible superradiant phases. In Section~\ref{sec:natureOfmode}, we examine the nature of these modes, showing that one of them exhibits an intricate decomposition in different bands due to the transverse pump-cavity angle deviating from $90^\circ$. By further exploring this angular dependence, we uncover a more general mechanism underlying the decomposition of  excitation modes. Finally, in Section~\ref{sec:dissipationInducedInstability}, we investigate the role of dissipation, demonstrating how it can synchronize the two excitation modes and ultimately lead to the emergence of exceptional points (EP), linking this work again to the companion paper~\cite{Natale2026Synchronization}.

\section{Bandstructure of a coupled BEC-cavity system}
\label{sec:system}
In this section, we introduce the system and establish the terminology for our analysis. We begin by describing the different light fields in our setup and the lattice structures they create in both real and reciprocal space. We then outline key concepts of band structure formalism, which will play a crucial role in our later analysis. Finally, we develop a mean-field model that predicts the qualitative phase diagram of the system and captures the softening of the excitation modes associated with the two superradiant phases.

\subsection{BEC-Cavity system}
We investigate a BEC trapped in the optical mode of a high-finesse cavity (see Fig.~\ref{fig:setup}(a)). The cavity mode forms an angle $\theta$ with the  transverse pump (TP). This TP is in the dispersive limit, i.e. its frequency $\omega_\mathrm{TP}$ is far detuned  by $\Delta_a=\omega_\mathrm{TP}-\omega_a$ from the atomic resonance $\omega_a$. It consists of two counter propagating running wave light fields with independent Rabi rates $\Omega_1$ and $\Omega_2$. We describe the imbalance in these beams with the parameter $\gamma=\sqrt{\Omega_1/\Omega_2}$. The cavity mode can only be populated with photons if these were scattered via the atoms from the TP into the cavity. If this process, which mediates the effective interactions between the atoms~\cite{Defenu2023Long}, is sufficiently strong, the quantum gas can overcome the kinetic energy cost of a density modulation and self-order in a spatial pattern efficiently scattering light into the cavity.

The single-particle Hamiltonian describing an atom in this setting reads ($\hbar=1$)
\begin{align}
    \mathcal{\hat{H}} =&  - \Delta_\mathrm{c} \hat{a}^\dagger \hat{a} + \frac{\mathbf{\hat{p}}^2}{2m} + \Vtp \cos^2(\kp\mathbf{x}) + U_0\hat{a}^\dagger \hat{a}\cos^2(\kc\mathbf{x}) \nonumber\\
    &+ \delta_+ \frac{\hat{a}+\hat{a}^\dagger}{2}\cos(\kp\mathbf{ x})\cos(\kc\mathbf{x})\nonumber \\ 
    &+ \delta_- \frac{\hat{a}-\hat{a}^\dagger}{2i}\sin(\kp\mathbf{ x})\cos(\kc\mathbf{x}),\label{eq:MainHamiltonian}
\end{align}
where $\hat{\mathbf{p}}$ is the momentum operator, $\hat{a}$ is the bosonic annihilation operator for a photon in the cavity mode, and  $\dc=\omega_\mathrm{TP}-\omega_c$ is the detuning of the TP frequency from the cavity resonance $\omega_c$. We find four periodic optical potentials acting on the atoms: $(i)$ The balanced part of the counter propagating TP forms a standing wave lattice along the pump wave vector $\kp$ with amplitude $\Vtp=\Omega_1\Omega_2/\Delta_a$. The imbalanced part is a simple running wave, leading to a constant offset which we neglect. $(ii)$ The light field in the cavity results in a second optical lattice along the cavity wave vector $\kc$ with amplitude $\hat{a}^\dagger \hat{a} U_0$, given by the number of photons $\hat{a}^\dagger \hat{a}$ in the cavity and  the rate $U_0=g^2/\Delta_a$ with  the vacuum Rabi coupling $g$. Here, $|\kp| = |\kc| = 2\pi / \lambda$, with the wave length $\lambda$ of the TP field. $(iii)$ The interference of the real part of the cavity field with the  balanced part of the TP forms a third potential (second line in Eq.~\eqref{eq:MainHamiltonian}) which has a checkerboard shape along $\kp$ and $\kc$.
$(iv)$ Finally, the fourth potential results from the interference of the running wave, imbalanced part of the TP with the imaginary part of the cavity field (third line in Eq.~\eqref{eq:MainHamiltonian}). Also this interference potential has a checkerboard structure, but it is shifted in space with respect to $(iii)$. Here, we have introduced $\delta_\pm:=\sqrt{\Vtp U_0}(\gamma\pm1/\gamma)$ as the strengths of the two interference potentials.

These four potential terms have two different periodicities: $(i)$ and $(ii)$ have a periodicity of $\lambda/2$, whereas the interference potentials $(iii)$ and $(iv)$ are $\lambda$-periodic. In order to describe our system, we choose a real space Wigner-Seitz (WS) cell, as shown in Fig.~\ref{fig:setup}(b) with size $\lambda/2$ along both $\mathbf{x}$ and $\mathbf{y}$. The intensity maxima of the TP and cavity lattices are indicated with yellow and blue lines, respectively. The WS cell encompasses one period of the TP and cavity lattice and half a period of the two interference terms. The Brillouin zone (BZ) is chosen accordingly and is spanned by the vectors $\kp$ and $\kc$. In Fig.~\ref{fig:setup}(c), two additional vectors are highlighted, pointing to the corners of the BZ, $\kpl:=\kp+\kc$ and $\kmi:=\kp-\kc$. These wave vectors are crucial for the underlying microscopic scattering processes, where a photon is scattered at an atom from the TP into the cavity, and the atom recoils either along $\kpl$ or $\kmi$.

\begin{figure}[ht]
\centering
\includegraphics[width=0.48\textwidth]{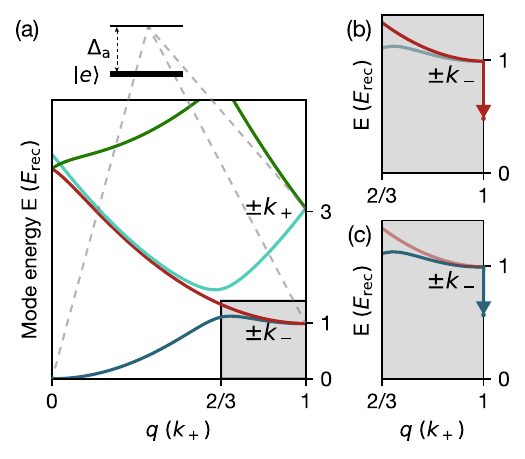}
\caption{\textbf{Band structure with coupling induced by the interference potential.} (a) Numerically calculated band structure along $\kpl$ for a potential with $\Vtp=1~\Erec$, $U_0\hat{a}^\dagger\hat{a}=\delta_\pm=0$ and a TP-cavity angle of $\theta=60^\circ$. The chosen axis points towards the corner of the Brillouin zone (BZ) at $\mathbf{q}=\kpl$. At this corner, four states are expected: $\pm\kpl$ and $\pm\kmi$ in the Bloch bands $(s,p,d,f)$, located at $3~\Erec$ and $1~\Erec$ respectively. This energy difference is caused purely by the system's geometry, not the by the applied potentials. The TP lattice induces a gap between two states, at $\mathbf{q}=2/3~\kpl$. The interference lattice mediates a two photon coupling process via the dispersively coupled excited atomic state ($\ket{e}$), linking the BEC at $\mathbf{q}=0$ with states located at $\mathbf{q}=\pm\mathbf{k}_\pm$. The detuning of the transverse pump relative to the atomic transition is given by $\Delta_a=\omega_\mathrm{TP}-\omega_a$. Panels (b) and (c) show a zoom-in on the gray-shaded area in the band structure. The coupling induced by the interference potential leads to a softening of two roton modes located at $\mathbf{q}=\pm\kmi$, as indicated by the arrows.}
\label{fig:bandcut}
\end{figure}

\subsection{Band structure formalism}
To study the BEC in these periodic potentials, we move to a momentum space description. Before applying it to our situation, we first highlight some important features of this band structure formalism, and consider a generic single-particle Hamiltonian of the form
\begin{align}
    \mathcal{H}=\frac{\mathbf{\hat{p}}^2}{2m}+V(\mathbf{x}),
\end{align}
with a periodic potential $V(\mathbf{x})=V(\mathbf{x}+\mathbf{a})$, employ Bloch's theorem and find eigenstates of the form
\begin{align}
    \phi_{\alpha,\mathbf{k}}(\mathbf{x})=\frac{1}{\sqrt{A}}\eikx \unkx,
\end{align}
where $\unkx$ is an $\mathbf{a}$-periodic function in real space, $\alpha$ denotes the band index, and $\mathbf{k}$ the momentum of the particle. We expand both the potential and the periodic function $\unkx$ in terms of reciprocal lattice vectors $\mathbf{G}$
\begin{align}
    V(\mathbf{x})&=\sum_\mathbf{G}V_\mathbf{G}e^{i\mathbf{G}\mathbf{x}},\\
    \unkx&=\sum_\mathbf{G}c_\mathbf{G}e^{-i\mathbf{G}\mathbf{x}}.\label{eq:blochExpansion}
\end{align}
This allows us to cast the Schrödinger equation for each $\mathbf{G}$ into reciprocal space, 
\begin{align}
    \frac{\hbar^2}{2m}(\mathbf{k}-\mathbf{G})^2 c_\mathbf{G}+\sum_\mathbf{G'}V_\mathbf{G'-G}c_\mathbf{G'}=\epsilon_{\alpha,\mathbf{k}}c_\mathbf{G}.
    \label{eq:reciprocalEV}
\end{align}
By numerically solving this set of linear equations for $c_\mathbf{G}$ in Eq.~\eqref{eq:reciprocalEV}, we find eigenvalues $\epsilon_{\alpha,\mathbf{k}}$, corresponding to the energy of a Bloch state at $\mathbf{q}=\mathbf{k}$, in the $\alpha$-th band. The corresponding eigenvector $c_\mathbf{G}$ can be used to recover the Bloch function according to Eq.~\eqref{eq:blochExpansion}.
The kinetic part of Eq.~\eqref{eq:reciprocalEV} is diagonal, whereas the potential mixes components of different reciprocal lattice vectors, which differ by the periodicity of the potential. More generally, a 1D potential with wave vector $k$ couples states with an integer multiple of $k$ distance in reciprocal space. This can also be seen from the calculation of the coupling between two Bloch states facilitated by a running wave $e^{ikx}$,
\begin{align}
    \bra{\phi_{n,k_l}}&e^{ikx}\ket{\phi_{m,k_m}}\nonumber\\
    &=\frac{1}{2}\int_{W.S.} e^{i(k_m-k_l+k)x}u_{n,k_l}^*(x)u_{n,k_m}(x)dx\nonumber\\
    &=\bra{u_{n,k_l}}\ket{u_{n,k_m}}\delta(k_m-k_l+k).
\end{align}

A $\cos(kx)$ potential thus couples states separated by $\pm k$ in reciprocal space. This can be generalized to two dimensions, where the two interference potentials in our system couple states spaced by $\Delta_k=\pm(\kp\pm\kc)$. For instance, the BEC state at $\mathbf{q}=0$ is coupled to the states at $\mathbf{q}=\pm\kpl$ and $\mathbf{q}=\pm\kmi$, as expected from the microscopic scattering picture described earlier. This process is illustrated in Fig.~\ref{fig:bandcut}(a) with gray dashed lines, representing the two-photon transition via the dispersively coupled excited state $\ket{e}$ of the atom. Due to the toroidal topology of a 2D Brillouin zone, the four states at $\pm\kmi$ and $\pm\kpl$, i.e. at the corners of the BZ, are at the same location in reciprocal space, although at different energies. We therefore use $\kmi$ and $\kpl$ interchangeably for the location of these states.

Fig.~\ref{fig:bandcut}(a) shows the band structure along $\kpl$ for the Hamiltonian in Eq.~\eqref{eq:MainHamiltonian}, with $\Vtp=1\,E_\mathrm{rec}$, zero cavity photons and a cavity-TP angle of $\theta=60^\circ$. This band structure thus only takes into account the presence of the 1D TP lattice, but is plotted along the coordinate  suited for analyzing the scattering process leading to a non-vanishing cavity field. It is obtained by numerically solving the eigenvalue problem given in Eq.~\eqref{eq:reciprocalEV} for the Hamiltonian $\mathcal{H}=\mathbf{\hat{p}}^2/2m + \Vtp \cos^2(\kp\mathbf{x})$. 

We focus on the four lowest Bloch bands $(s,p,d,f)$, which -- at the corners of the BZ -- correspond to the four states found at $\pm\kmi$ and $\pm\kpl$. The TP lattice opens a gap between energetically degenerate states, spaced by $2\kp$ in reciprocal space. Along the chosen direction $\kpl$, this results in an avoided crossing at $\mathbf{q}=\csc(\theta)\,\kpl/\abs{\kpl}$ for the two bands ($s$ and $d$). For $\theta=60^\circ$, this results in a band gap at $\mathbf{q}=2/3~\kpl$. Meanwhile, the  states at the corners of the BZ remain to lowest order unaffected by the TP potential. The gap between the $\kmi$ and $\kpl$ states appearing at $1~E_\mathrm{rec}$ and $3~E_\mathrm{rec}$ arises purely from our choice of geometry. Appendix \ref{ap:overlapPlots} presents the energy scaling of these four states with respect to $\Vtp$.

\subsection{Mean field phase diagram}
\label{subsec:MFPD}
We are now in the position to  introduce a mean-field treatment of our system, enabling us to predict the softening of the excitation modes and to construct the phase diagram of the system. Our approach is based on the following hierarchy of potential terms: we assume that the TP potential $\Vtp$ dominates over the cavity potential $U_0\hat{a}^\dagger\hat{a}$. Consequently, we diagonalize the TP potential using the band structure formalism as before, treat the interference potentials ($\propto\sqrt{\Vtp U_0\hat{a}^\dagger\hat{a}}$) as a perturbation on the resulting Bloch states, and neglect the cavity potential $\propto U_0\hat{a}^\dagger\hat{a}$. 
This approximation holds outside of the superradiant phase, where the cavity field is negligible. Inside the self-ordered phases, the same hierarchy persists, meaning our results remain qualitatively valid. However, especially for large $\Vtp$, the presence of the cavity potential is expected to introduce quantitative corrections.

We start by making a few-momentum mode ansatz for the many-body atomic wavefunction
\begin{align}
    \hat{\Psi}(\mathbf{x}) = \frac{1}{\sqrt{A}} \left( \phi_{s, \mathbf{q}=0}(\mathbf{x})\hat{c}_0 + \sum_{(\alpha,\mathbf{q'})}\phi_{\alpha,\mathbf{q'}}(\mathbf{x})\hat{c}_{\alpha,\mathbf{q'}}\right),
\end{align}
where the index $\alpha$ runs over the four lowest bands $\alpha\in\{s,p,d,f\}$ and $\mathbf{q'}\in\{\kmi,\kpl\}$.
This wavefunction is normalized to the Wigner-Seitz unit cell and is composed of Bloch functions $\phi_{\alpha,\mathbf{q}}(\mathbf{x})$ for a given quasi momentum $\mathbf{q}$ and band index $\alpha$. The $\hat{c}_{\alpha,\mathbf{q}}$ are bosonic operators, annihilating a phononic particle with wave vector $\mathbf{q}$ in the given band $\alpha$.

We use the Hamiltonian in Eq.~\eqref{eq:MainHamiltonian}, neglect the cavity lattice, integrate the ansatz over the Wigner-Seitz cell, and find the following reduced Hamiltonian:
\begin{align}
    \hat{H}=& \sum_i \omega_i(\Vtp) \hat{c}_i^\dagger\hat{c}_i- \Delta_\mathrm{c} \hat{a}^\dagger \hat{a} \nonumber\\
    &+ \delta_+ \frac{\hat{a}+\hat{a}^\dagger}{2} \sum_{i,j} \Theta_{i,j}(\Vtp) \hat{c}^\dagger_i\hat{c}_j\nonumber \\ 
    &+ \delta_- \frac{\hat{a}-\hat{a}^\dagger}{2i}\sum_{i,j} \Xi_{i,j}(\Vtp) \hat{c}^\dagger_i\hat{c}_j,\label{eq:Hreduced}
\end{align}
with the overlap integrals
\begin{align}
    \Theta_{i,j} &:= \frac{1}{A}\int_{W.S.}\phi^*_{i}(\mathbf{x})\cos(\kp\mathbf{x})\cos(\kc\mathbf{x})\phi_{j}(\mathbf{x})\mathbf{dx},\\
    \Xi_{i,j} &:= \frac{1}{A}\int_{W.S.}\phi^*_{i}(\mathbf{x})\sin(\kp\mathbf{x})\cos(\kc\mathbf{x})\phi_{j}(\mathbf{x})\mathbf{dx}.
\end{align}
The indices $i$ and $j$ enumerate all band indices $\alpha$ and quasi momenta $\mathbf{q}$ used in the ansatz. The energy $\omega_i$ and the overlap integrals depend on $\Vtp$, since the Bloch functions depend on $\Vtp$ (which for clarity is not stated explicitly). Note, that $\Theta_{i,j}=\Theta_{j,i}^*$ and analogous for $\Xi$. 

We calculate these overlap integrals numerically by computing the Bloch wavefunctions for the given momenta and a fixed TP lattice depth. A plot of the overlaps as a function of $\Vtp$ is provided in appendix \ref{ap:overlapPlots}. 
Our results show that all overlap matrix elements vanish except for $\Theta_{i,0}$ with $i\in\{(s,\kmi),(f,\kpl)\}$ and for $\Xi_{i,0}$ with $i\in\{(p,\kmi),(d,\kpl)\}$, along with their complex conjugates. This implies that there is no direct coupling among the four Bloch states at $\mathbf{q}=\mathbf{k}_\mathrm{\pm}$; instead, each of these states couples only to the BEC state with $i=(s,\mathbf{0})$.

We exploit the separation of time scales in our system to find an atom-only description: the atomic motion is governed by $\Erec$, which evolves on a time scale much slower than cavity dissipation characterized by $\kappa$. We thus can adiabatically eliminate the cavity field by imposing
\begin{align}
    \frac{\mathrm{d} \hat{a}}{\mathrm{d} t} &= i[\hat{H},\hat{a}] - \kappa \hat{a} = 0.
\end{align}
This allows us to write the cavity field as a function of the atomic density modulation,
\begin{align}
    \hat{a} =& \frac{\kappa + i \dc}{2(\dc^2+\kappa^2)}\sum_{i,j}\left(\delta_-\Xi_{i,j} \hat{c}^\dagger_i\hat{c}_j-i\delta_+\Theta_{i,j} \hat{c}^\dagger_i\hat{c}_j\right).
\end{align}
Using this expression for $\hat{a}$ and Eq.~\eqref{eq:Hreduced}, we find the effective atom-only Hamiltonian:
\begin{align}
    &\hat{H}_\mathrm{eff}= \sum_i \omega_i(\Vtp) \hat{c}_i^\dagger\hat{c}_i +\frac{\dc}{2(\dc^2+\kappa^2)}\nonumber\\
    &\times\left[\left(\delta_-\sum_{i,j}\Xi_{i,j} \hat{c}^\dagger_i\hat{c}_j\right)^2+ \left(\delta_+\sum_{i,j}\Theta_{i,j}\hat{c}^\dagger_i\hat{c}_j\right)^2\right].\label{eq:Heffektiv}
\end{align}

We move to a mean-field description, where $c_i=\expval{\hat{c_i}}$.
To determine the mode decomposition of the excited states and their energies around an equilibrium point, we perform a quadratic expansion around the system's ground state. Specifically, we compute the Bogoliubov matrix of the effective Hamiltonian and evaluate it at $c_0=\sqrt{N}$ and $c_i=0$. The full expression for the Bogoliubov matrix is provided in appendix \ref{ap:FreeEnergyFunctional}. Notably, the Bogoliubov matrix is not initially in diagonal form and must be diagonalized. Its eigenvalues correspond to the energies of the system’s excitation modes, while the eigenvectors reveal their decomposition into different Bloch functions (or the corresponding $\mathbf{k}_\pm$ momentum contributions). The couplings induced by the two interference potentials thus lead to the independent softening and mixture of the four states at the corners of the BZ, as indicated in Fig.~\ref{fig:bandcut}(b) and (c) by the vertical arrows. While all four states are coupled to the BEC state and evolve in energy depending on system parameters, we find that only the two states originating at $\SI{1}{\Er}$ (i.e. the two $\kmi$ states) can soften fully to zero energy. We therefore focus on these two modes which are sufficient to predict the phase diagram.

\begin{figure}[ht]
\centering
\includegraphics[width=0.48\textwidth]{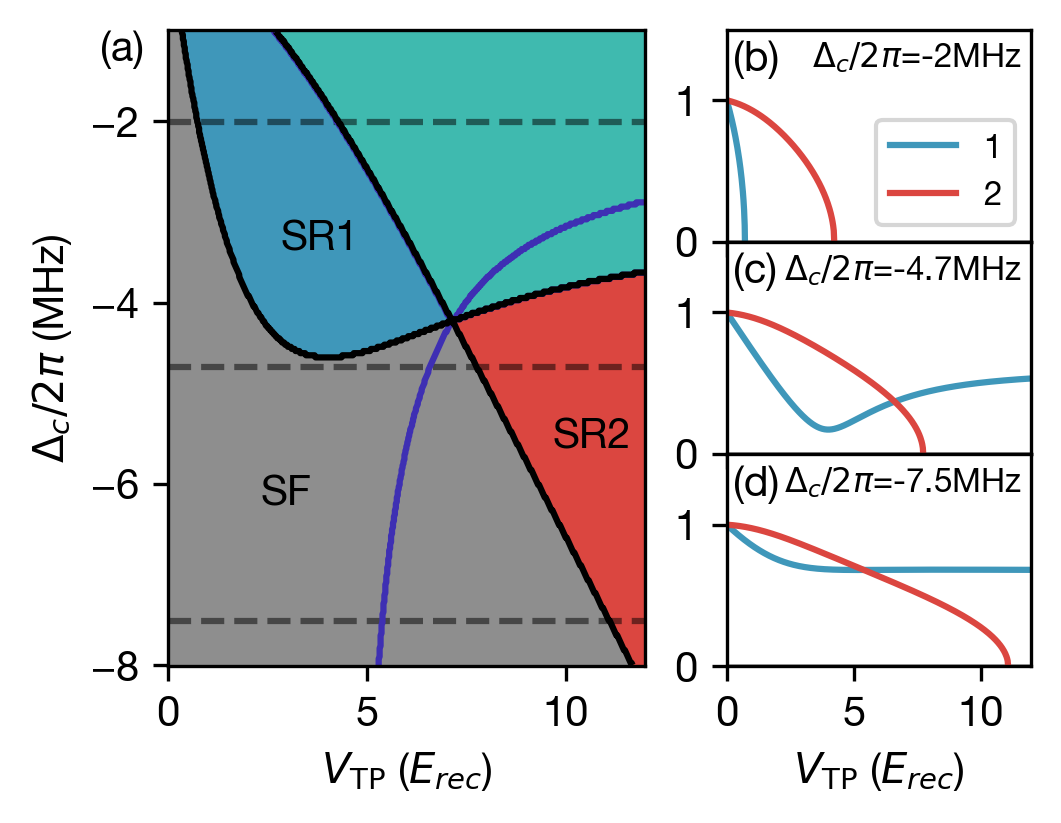}
\caption{\textbf{Mean-field phase diagram} (a) Phase diagram obtained from a quadratic expansion of the free-energy functional, plotted as a function of $\dc$ and $\Vtp$. The gray region corresponds to the superfluid phase (SF), where no self-organization occurs. The two coloured regions (red and blue) indicate the parameter ranges where the system crystallizes into one of two superradiant density patterns (SR1 or SR2). The black solid lines mark the points where an excited mode reaches zero energy. Between SF and SR this signals the occurrence of a phase transition. In the turquoise region both modes soften below zero energy. The blue line represents the parameter values where the two roton modes become degenerate in energy. Panels (b)-(d) show the mode energies along horizontal cuts through the phase diagram (indicated by gray dashed lines in (a)) for different cavity detunings $\dc/2\pi=-2,-4.7,-7.5~\mathrm{MHz}$. The two modes exhibit different softening behaviour: while the SR2 mode always softens for sufficiently large $\Vtp$ independent of $\dc$, the precursor mode of SR1 only softens up to a certain $\Vtp$ value, before becoming harder again.}
\label{fig:MFPD}
\end{figure}

We calculate the energy of these two excited modes across the parameter space defined by the TP lattice depth $\Vtp$ and cavity detuning $\dc$. Tracking the energies of both excitation modes starting at $\SI{1}{\Er}$, we identify the points where they reach zero, signalling a phase transition. Figure~\ref{fig:MFPD}(a) presents the resulting phase diagram for a blue-detuned TP. The gray region corresponds to the superfluid phase (SF), while the blue and red regions indicate two  distinct superradiant phases (labelled SR1 and SR2), reached when one of the two modes has fully softened. We thus identify the corresponding excited modes as the precursors of the two phases SR1 and SR2, and from now on refer to them as \textit{SR1 mode} and \textit{SR2 mode}, respectively. In the turquoise region, both excitation modes have softened below zero energy. The blue line marks the points where the two modes become degenerate in energy. Above this line, the system is expected to enter the SR1 phase if it self-orders, while below it, SR2 should dominate. However, as shown experimentally in Ref.~\cite{li2021}, the transition between these two phases is first-order, meaning that the occupied phase near the blue line depends on the system’s history.

Figures~\ref{fig:MFPD}(b)-(d) show the energies of the two excitation modes as a function of $\Vtp$ for selected cavity detunings, indicated by the gray dashed lines in Fig.~\ref{fig:MFPD}(a). These plots illustrate that, depending on the cavity detuning, both modes can soften independently to zero. The softening process arises from a competition between the atom's potential and kinetic energy, i.e. the coupling induced by the interference lattice versus the band gap opening due to increasing TP power -- see Equation~(\ref{eq:Heffektiv}). This behaviour is governed by the  overlap integrals $\Xi_{i,0}(\Vtp)$ and $\Theta_{i,0}(\Vtp)$ which characterize the coupling between the BEC state and the excited states, as shown in Appendix~\ref{ap:overlapPlots}.
The SR1 mode depends on $\Theta_{i,0}$, which decreases with increasing $\Vtp$. In addition, the dominant $f$ band shifts upward in energy as $\Vtp$ increases. As a result, the SR1 phase exhibits reentrant behavior, meaning that beyond a certain TP power, the system exits this phase. In contrast, the SR2 mode is governed by $\Xi_{i,0}$, which increases monotonically with $\Vtp$. Moreover, the dominant $p$ band decreases in energy with increasing $\Vtp$. These combined effects ensure that once the system enters the SR2 phase, it remains there as TP power continues to rise.

\section{Nature of the two SR modes}
\label{sec:natureOfmode}
\subsection{Mode decomposition}

We analyse the decomposition of the two precursor modes into different Bloch bands and real space patterns. We project the eigenstates for the SR1 mode and the SR2 mode derived from the quadratic expansion back onto the original basis of the four Bloch states in their corresponding bands $\alpha\in\{s,p,d,f\}$. 

\begin{figure}[ht]
\centering
\includegraphics[width=0.48\textwidth]{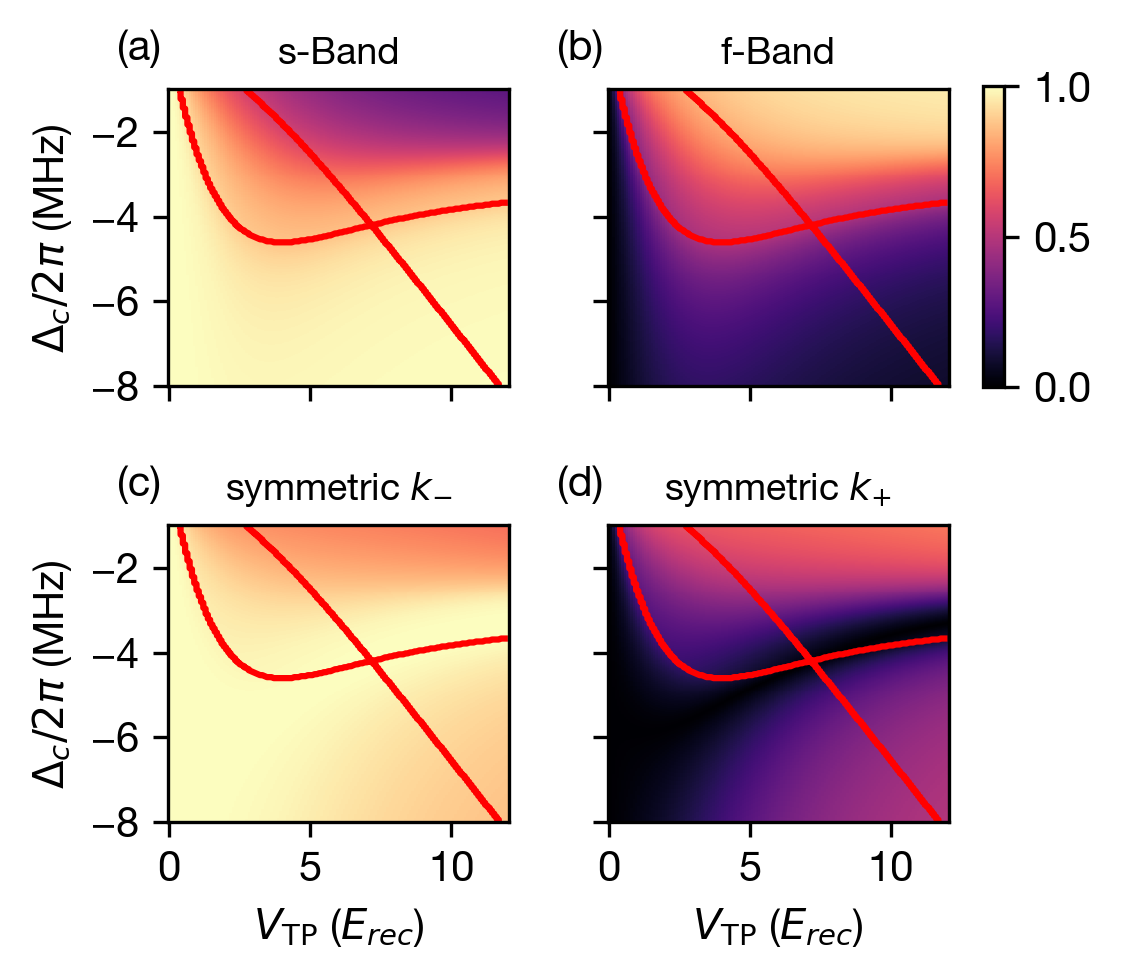}
\caption{\textbf{Symmetry decomposition of the precursor mode of SR1.} The red lines indicate the crossing of an excited mode with zero energy, as determined in the previous mean-field analysis. Panels (a) and (b) show the projection of the SR1 mode onto the two Bloch states coupled by this phase. For low $\Vtp$, the SR1 mode consists predominantly of $s$ band contributions, for larger $\Vtp$ and small negative $\dc$ more of the $f$ band is admixed. Panels (c) and (d) show the projection of the Bloch states on a symmetric superposition of $\kmi$ and $\kpl$, giving insight into the real space density pattern. Note, that this basis truncates the contribution of higher $\mathbf{k}$ modes and is incomplete for large $\Vtp$. 
}
\label{fig:decomp1}
\end{figure}
In Fig.~\ref{fig:decomp1}, we show the detailed decomposition of the SR1 mode. Only the $s$ and the $f$ band contribute (shown in panels (a) and (b)), while the $p$ and the $d$ band are not involved. This can be understood from the difference in the symmetries of the two interference potential terms, as mentioned before. The $s$ band state at $\kmi$ is the antisymmetric superposition of $\cos(\kpl)$ and $\cos(\kmi)$, whereas the $f$ band state is the symmetric superposition of all four momentum components, i.e. $\beta_1\cos(\kpl)+\beta_2\cos(\kmi)$ ($\beta_i$ are real positive coefficients). Note, that the Bloch states themselves are changing for different values of $\Vtp$, i.e. the contribution of $\kmi$ and $\kpl$. We show in Fig.~\ref{fig:decomp1}(c) and (d) the projection of the SR1 mode onto the symmetric superposition of $\kmi$ and $\kpl$. This is, compared to the Bloch states not a complete basis, as for high $\Vtp$ higher order momentum states are occupied. However, in the displayed regime it allows us to understand the real space decomposition of the mode, as we show below.

\begin{figure}[ht]
\centering
\includegraphics[width=0.48\textwidth]{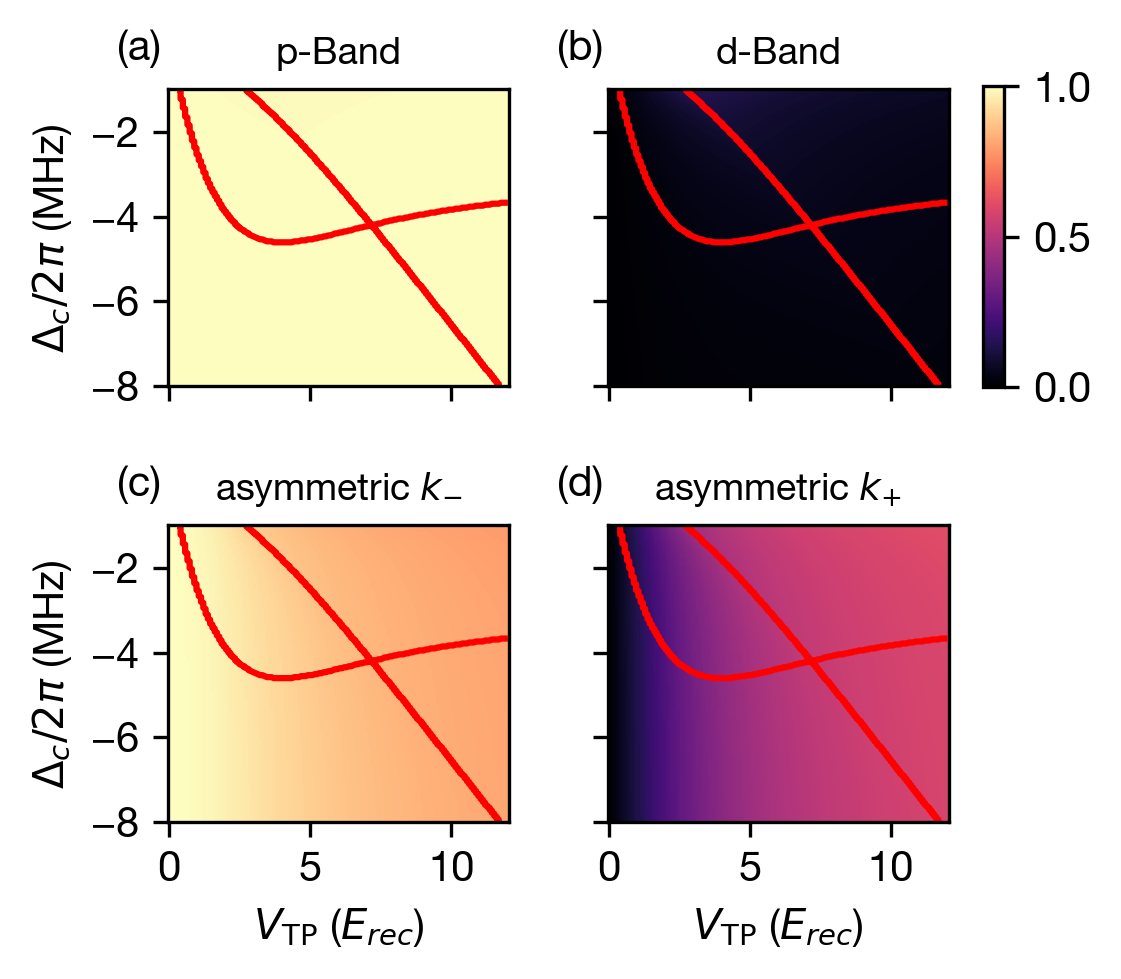}
\caption{\textbf{Symmetry decomposition of the precursor mode of SR2.} The red lines indicate the zero-energy crossing of an excited mode as determined in our previous analysis. Panels (a) and (b) display the decomposition of the SR2 mode into the two coupled Bloch states, the $p$ and the $d$ band. The SR2 mode is purely determined by the $p$ band. Panels (c) and (d) indicate that the $p$ band has a real space distribution of a 1D asymmetric superposition of $\pm\kmi$ for low $\Vtp$ and starts to admix the asymmetric superposition of $\pm\kpl$, turning it into a 2D checkerboard pattern. }
\label{fig:decomp2}
\end{figure}

For the SR2 mode the situation is more straight forward (see Fig.~\ref{fig:decomp2}). This mode has only a non-zero contribution from the $p$ band. The projection onto the asymmetric superposition of $\kmi$ and $\kpl$ therefore only reflects the decomposition of the $p$ band Bloch state.

\subsection{Real-space representation}

\begin{figure}[h!]
\centering
\includegraphics[width=0.48\textwidth]{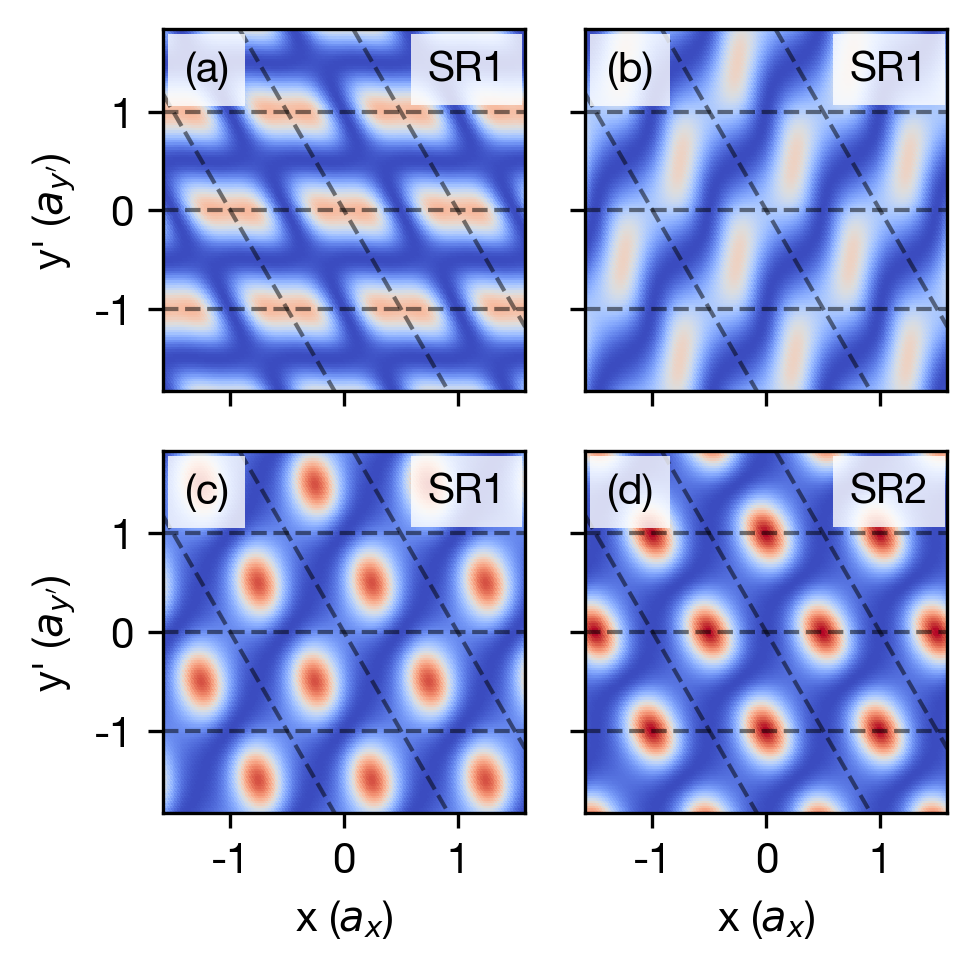}
\caption{\textbf{Real space representation of the precursor mode.} The axis are along $\mathbf{x}$ in units of the lattice spacing $a_x$ and orthogonal to this, which does not correspond to a natural axis of the system. The dashed lines represent the maxima of the transverse pump ($\Vtp$) (diagonal lines) and the cavity lattice (horizontal lines). Both wave functions are normalized to the Wigner-Seitz unit cell. Panels (a)-(c) show the real space plot of the precursor mode corresponding to SR1 at $\Vtp=4~E_\mathrm{rec}$ and at $\dc/2\pi=-1,-4.5,-10~\mathrm{MHz}$ respectively. Panel (d) shows the real space representation of the precursor mode of SR2 at $\Vtp=4~E_\mathrm{rec}$ and at $\dc/2\pi=-4.5~\mathrm{MHz}$. In contrast to SR1, this mode exhibits a pure checkerboard pattern with similar modulation depth in $\kpl$ and $\kmi$. Here, the atomic wavefunction has its maxima at the minima of the TP lattice but at the maxima of the cavity lattice.}
\label{fig:realspace}
\end{figure}

These decompositions can be translated into real space representations of the wavefunction. We show both modes in real space for different points in the phase diagram in Fig.~\ref{fig:realspace}. The real space plots are taken at $\Vtp=4~E_\mathrm{rec}$ and at $\dc/2\pi=-1,-4.5,-10~\mathrm{MHz}$, for SR1 from (a) to (c). As the mode of SR2 does not change significantly with $\dc$, we show it only at $\dc/2\pi=-4.5~\mathrm{MHz}$.  The gray dashed lines indicate the maxima of the cavity and the TP lattice, horizontal and diagonal, respectively. Note, that we expect the ground state in the self-ordered phase to be a superposition of the SF state and the excited modes. We also expect a quantitative modification of this result for a strong cavity field, occurring at high $\Vtp$, while the qualitative structure remains.

The spatial character of the SR1 mode changes as a function of the detuning. We start by looking at the situation with $\dc/2\pi=-1~\mathrm{MHz}$ (panel (a)), which is well inside the self-ordered phase SR1. The atomic wave function of the SR1 mode forms a checkboard pattern with the occupation maxima coinciding with the TP and cavity lattice maxima, thus maximizing the scattering of photons from the TP into the cavity mode. Moving to $\dc=-4.5~\mathrm{MHz}$ (panel (b)), i.e. at the lower edge of the SR1 phase, we find an almost pure 1D density modulation in $\kmi$ direction. This wave function still has a finite overlap with the TP lattice and is therefore able to scatter light into the cavity, allowing it to soften to the point of a phase transition. Finally, for $\dc/2\pi=-10~\mathrm{MHz}$, meaning well outside the self-ordered phases, the wave function of the SR1 mode forms again a checkerboard pattern, however located at the minima of both lattices, thus not able to trigger a phase transition.
This behaviour can also be understood by going back to the decompositions discussed in the previous section. If we investigate SR1 in terms of a density modulation along $\kmi$ and $\kpl$, we find that the SR1 is predominantly modulated along $\kmi$ for low $\Vtp$ and for an arch in parameter space visible as the black area in Fig.~\ref{fig:decomp1}(d). Above and below this arch, the density modulation is more balanced between $\kmi$ and $\kpl$ direction, leading to the checkerboard nature. This destructive interference of the $\kpl$ modulation can be understood from the admixture of the $f$ band, which carries a different sign between the two modulation directions compared to the $s$ band. How strong this cancellation is depends on the precise mixing ratios as well as on the decomposition of the Bloch states themselves. The physical origin of this effect will be discussed in the subsequent subsection~\ref{subsec:angularDependence}.

The nature of the SR2 mode is again simpler. As shown in Fig.~\ref{fig:decomp2}, a density modulation in $\kmi$ direction dominates for low $\Vtp$. As $\Vtp$ is increased, the higher order mode of $\kpl$ starts to admix and results in a density modulation along both $\kpl$ and $\kmi$, i.e. a checkerboard. As shown in Fig.~\ref{fig:realspace}(d), this checkerboard pattern exhibits occupation maxima on the cavity lattice maxima and the TP lattice minima. This pattern can still scatter light from the TP even though it has minimal overlap with the TP lattice. This is enabled by the imbalanced TP configuration, which results in a running wave of the TP having overlap with the atoms even when they occupy the minima of the TP lattice.
We note that the two density patterns of SR1 and SR2 are shifted by half a lattice site for all $\dc$, which agrees with the experimental observation, that the two modes couple to different quadratures of the cavity light field and that there is a phase shift of $\pi/2$ of this field when going from one phase to the other~\cite{zupancic2019}.

\subsection{Angular dependence}
\label{subsec:angularDependence}
\begin{figure*}[ht]
\centering
\includegraphics[width=0.95\textwidth]{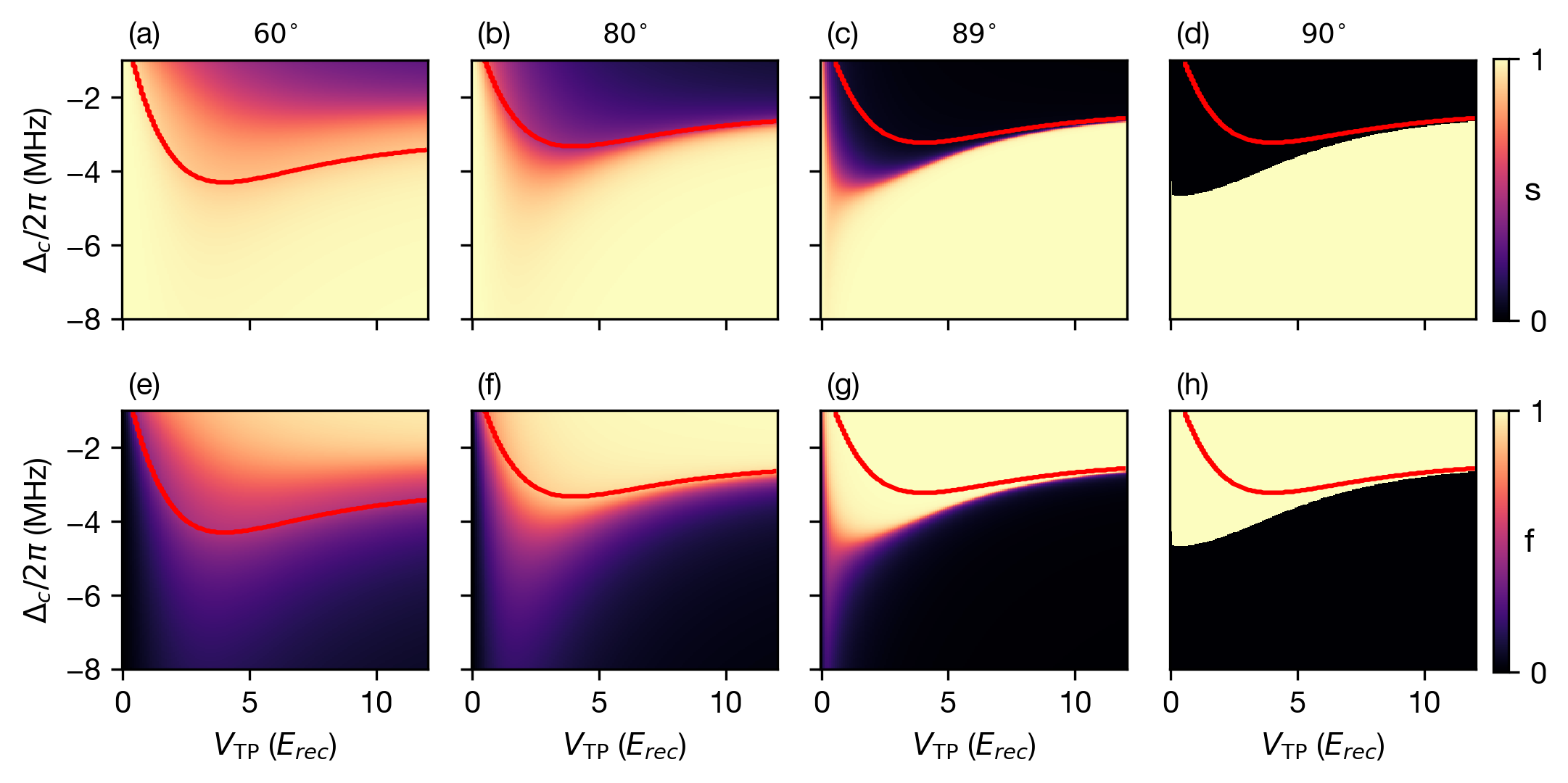}
\caption{\textbf{SR1-mode nature as a function of the TP-cavity angle} In the balanced transverse pump (TP) case, we find only the SR1 phase in the phase diagram (with the phase boundary indicated by the red line). For each angle, we show the $s$ band (panels (a)-(d)) and $f$ band (panels (e)-(f)) occupation of the mode. Notably, the SR1 precursor mode has no contribution from the other two Bloch bands. \textbf{$\mathbf{90^\circ}$}, panels (d) and (h): At this angle, the $\kpl$ and $\kmi$ modes are degenerate, and we find that the mode responsible for SR1 forms a pure checkerboard pattern with $\coco$ symmetry. \textbf{$\mathbf{89^\circ}$}, panels (c) and (g): Shifting the angle by just $1^\circ$ introduces an asymmetry between $\kpl$ and $\kmi$, which starts to affect the decomposition outside of the phase. \textbf{$\mathbf{80^\circ}$} and \textbf{$\mathbf{60^\circ}$}, panels (a),(e),(b) and (f): As we move to steeper angles, the admixture of the $s$ band increases and becomes significant close to and beyond the phase transition. This is caused by the energy competition, between the $f$ band which drives self-organization, and the increased energy cost of $\kpl$ due to angular imbalance. The additional density modulation given by the $s$ band, does not directly contribute to the scattering process leading to superradiance, but reduces the energy cost by partially cancelling out the $\kpl$ density modulation.}
\label{fig:angularDepDecomp}
\end{figure*}

In this section, we want to highlight the significance of the angle between cavity and transverse pump. This generalisation allows to connect our findings to different experimental settings and will uncover the physical origin of the intricate decomposition of SR1. It is thus sufficient to consider the case of a balanced TP ($\gamma=1$, absence of SR2) for different angles. The results of this analysis are presented in Fig.~\ref{fig:angularDepDecomp}. Since the SR1 mode can only occupy the $s$ and $f$ band, we focus on these two components for each angle. The red line again indicates the phase transition, which weakly depends on the angle.

At an angle of 90$^\circ$, the mode dominating in and around the SR1 phase consists purely of the $f$ band. This is expected, as the atomic wave function maximises its overlap with the transverse pump in this configuration, enabling it to scatter light into the cavity. Additionally, at 90$^\circ$, all four bands are degenerate in energy -- meaning that density modulations along $\mathbf{k_+}$ and $\mathbf{k_-}$ have the same energy cost. As a result the system favours the mode most beneficial for self-organization.

However, as the angle deviates from 90$^\circ$, the energy cost of the two modulations along $\mathbf{k_+}$ and $\mathbf{k_-}$ become different. According to our definition of $\mathbf{k_\pm}$, the cost of $\mathbf{k_+}$ increases relative to $\mathbf{k_-}$. This imbalance creates a competition between the energy gain from self-organization and the energy penalty of occupying the $\mathbf{k_+}$ mode. To reduce this cost, the system admixes a checkerboard pattern with opposite sign on the $\mathbf{k_+}$ modulation, effectively suppressing it. While this additional density pattern costs kinetic energy, it does not contribute significantly to the scattering of light and the build-up of the cavity potential. We therefore expect this admixing to vanish when going to a deep TP lattice, where kinetic energy is not the dominant energy scale. Indeed, in Fig.~\ref{fig:angularDepDecomp} at 60 degrees, increasing the TP power $\Vtp$ leads to the $f$ band becoming the dominant component inside the phase SR1.

An additional competition arises for different values of $\dc$. From Eq.~\eqref{eq:Heffektiv} we find that the coupling terms scale proportional to $1/\dc$. For large negative $\dc$ these terms become small compared to the Bloch state energy. Starting from large negative  $\dc$ and decreasing in magnitude, the system approaches a regime where the kinetic energy competes with the coupling term and for small absolute $\dc$ the overlap terms dominate. Since the $f$ band has significantly stronger overlap, it dominates in this regime. 
These findings align with our study of the real-space density pattern of the excitation modes for $60^\circ$. Fixing $\Vtp$ and varying $\dc$, we find that for large absolute $\dc$ only the $s$ band is occupied, resulting in a checkerboard pattern. As $\dc$ approaches zero, the system transitions through the intermediate regime with a 1D density pattern. At small negative $\dc$, the checkerboard pattern re-emerges -- now dominated by the $f$ band and resulting in a spatial shift.

\section{Dissipation induced instability}
\label{sec:dissipationInducedInstability}
We have found in subsection~\ref{subsec:MFPD} how the two modes corresponding to SR1 and SR2 can cross in energy, without influencing each other. For the derivation of this result we have used an effective hermitian Hamiltonian and neglected the non-hermitian effect the cavity decay has on the system. In this section, we will explore the physics induced by this dissipative coupling between the two excitation modes. We will recover a description of the mode coalescence which is experimentally observed and discussed in the companion paper~\cite{Natale2026Synchronization}. We move from the quadratic expansion of the effective Hamiltonian as used in subsection~\ref{subsec:MFPD}, to a linearized dynamical description of the system. In particular, we will be interested in the mode crossing of the two excited modes occurring for certain parameter values.

\subsection{Dynamical model}
We start the discussion with the same few-mode expansion and integration of the real-space Wigner-Seitz cell as used in subsection~\ref{subsec:MFPD}, i.e. the Hamiltonian given in Eq.~\eqref{eq:Hreduced}. The dynamical equations for the atomic operators $\hat{c}_i$ ($0\leq i\leq 4$) and the photonic field $\hat{a}$ including the cavity loss $\kappa$ are given by
\begin{align}
    \frac{d}{dt}{\hat{c}_i}=&-i\omega_i\hat{c}_i\nonumber\\
    &-i\sum_{j}\left(\delta_+ \Re(\hat{a}) \Theta_{i,j}\hat{c}_j+\delta_-\Im(\hat{a}) \Xi_{i,j}\hat{c}_j\right),\\
    \frac{d}{dt}{\hat{a}}=&i\dc\hat{a}-\kappa \hat{a} \nonumber\\ 
    &- i\frac{\delta_+}{2} \sum_{i,j} \Theta_{i,j}\hat{c}_i^\dagger\hat{c}_j+ \frac{\delta_-}{2} \sum_{i,j} \Xi_{i,j}\hat{c}_i^\dagger\hat{c}_j \,.\label{eq:dynamicalEqFirstorder}
\end{align}

In order to solve this system of equations we assume the limit of large cavity dissipation and eliminate the dynamics of the cavity field as before. We again make a mean-field assumption, which allows us to write the operators as c-numbers $c_i\approx\langle\hat{c_i}\rangle$. Lastly we assume that the BEC (described by $c_0$) is macroscopically occupied i.e. $c_0\propto \sqrt{N}$. We therefore treat the occupation of the excited states only perturbatively, and derive the following set of equations
\begin{align}
    \frac{d^2}{dt^2}c_i=&-\omega_i^2c_i-\frac{N\dc\omega_i}{\dc^2+\kappa^2}\left(\delta_+^2\Theta_{0,i}^2+\delta_-^2\Xi_{0,i}^2\right)\Re(c_i)\nonumber\\
    &-\frac{N\omega_i}{\dc^2+\kappa^2}\sum_{j\neq i}\left[\delta_+\Theta_{0,j}\left(\dc\delta_+\Theta_{0,i}-\kappa\delta_-\Xi_{0,i}\right)\right.\nonumber\\
    &+\left.\delta_-\Xi_{0,j}\left(\kappa\delta_+\Theta_{0,i}+\dc\delta_-\Xi_{0,i}\right)\right]\Re(c_j)\,.
    \label{eq:dynamicalEqSecondorder}
\end{align}
Different from the hermitian case (cf. Equation~\ref{eq:Heffektiv}), the overlaps $\Theta_{i,j}$ and $\Xi_{i,j}$ here appear as products due to the non-zero cavity dissipation rate $\kappa$. The dissipation thus effectively leads to a coupling of the excitation mode SR1 and the excitation mode SR2. By diagonalizing this set of coupled equations, we obtain the potentially complex eigenvalues governing the system's evolution. Considering the dissipative nature of our system, we identify the real part of these eigenvalues with the energies of the excited modes in the parameter space.

\begin{figure}[ht]
\centering
\includegraphics[width=0.48\textwidth]{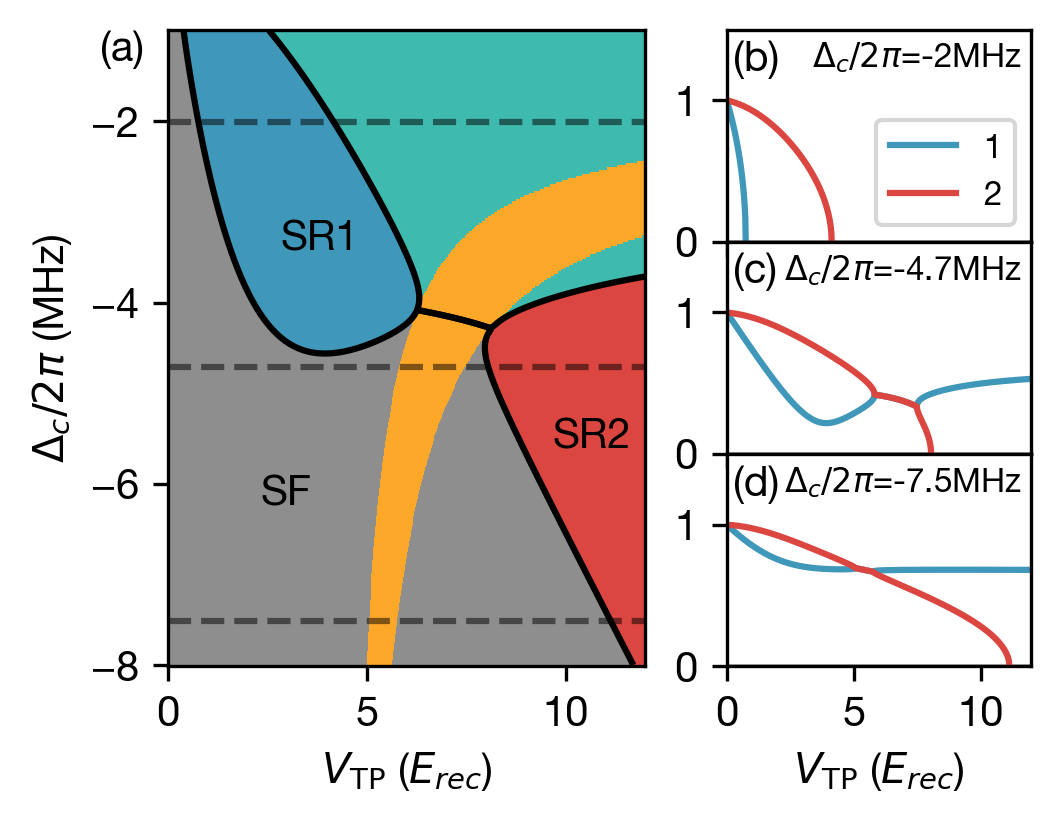}
\caption{\textbf{Mean-field phase diagram including the effect of dissipation} (a) The phase diagram as obtained from diagonalizing the dynamical equations, plotted for $\dc$ and $\Vtp$. The gray area indicates the system outside of the self-organized phase, labelled as SF. The black lines show where one of the two mode energies crosses zero and between SF and SR mark the expected phase transition. The blue area corresponds to SR1 and the red to SR2. The new feature compared to the previous mean-field analysis is the coalescence region of the two precursor modes, indicated by the orange area. Note that the coalescence is also predicted to occur within the phase; however, as this formalism is not self-consistent, we expect this region to be modified when a full self-consistent treatment is applied. Panels (b)-(d) show the real part of the mode energies along cuts through the phase diagram (gray dashed lines) for different cavity detunings $\dc$. For small negative $\dc$ the dissipative coupling starts to dominate when the energies of the two modes are degenerate. This happens at large $\Vtp$ for small negative $\dc$. For larger negative values $\dc$, the two modes intersect and coalesce as seen in panels (c) and (d).}
\label{fig:dissipativePD}
\end{figure}

We find a modified phase diagram, with similar features as the phase diagram obtained from the quadratic expansion of the free energy functional, see Fig.~\ref{fig:dissipativePD}(a). The two ordered phases SR1 and SR2 qualitatively retain their nature and shape. However, there is an additional area indicated in orange, which signifies the appearance of a non-zero complex part of the mode energy.  The location of this is centred around the blue line in Fig.~\ref{fig:MFPD}, where the energies of the two modes crossed. We can see in the different cuts (Fig.~\ref{fig:dissipativePD}(b)-(d)), that the real part of the two mode energies coalesce for a finite parameter range, instead of crossing as predicted by our previous analysis. This finding is experimentally confirmed in our companion paper~\cite{Natale2026Synchronization}. In the coalescence area the two modes evolve with the same energy, and will synchronize in phase. The energies of these modes becoming complex signifies the presence of an exponentially growing and an exponentially decreasing mode population, if evolved in time. As our system treats the occupation of these modes only perturbatively, we expect that we can only describe the initial escape rate of the system to such a mode dictated by the imaginary eigenvalue.

We observe this mode synchronization also inside of the superradiant phases, where the synchronizing modes SR1 and SR2 are both below the energy of the superfluid state and therefore occupied. However, as our formalism does not take the cavity field self-consistently into account, we expect the shape of this area to be quantitatively modified. We can still connect the appearance of a dynamical state between the two self-organized crystals to the dissipation-induced atomic motion, investigated in Ref.~\cite{dreon2022}.

\subsection{Mode decomposition with dissipation}
As discussed in the previous section, the presence of dissipation leads to a coupling between the two excitation modes. It is therefore interesting to revisit the decomposition of the SR1 and the SR2 modes into Bloch bands for the non-hermitian case, specifically in the coalescence region. 

\begin{figure}[ht]
\centering
\includegraphics[width=0.48\textwidth]{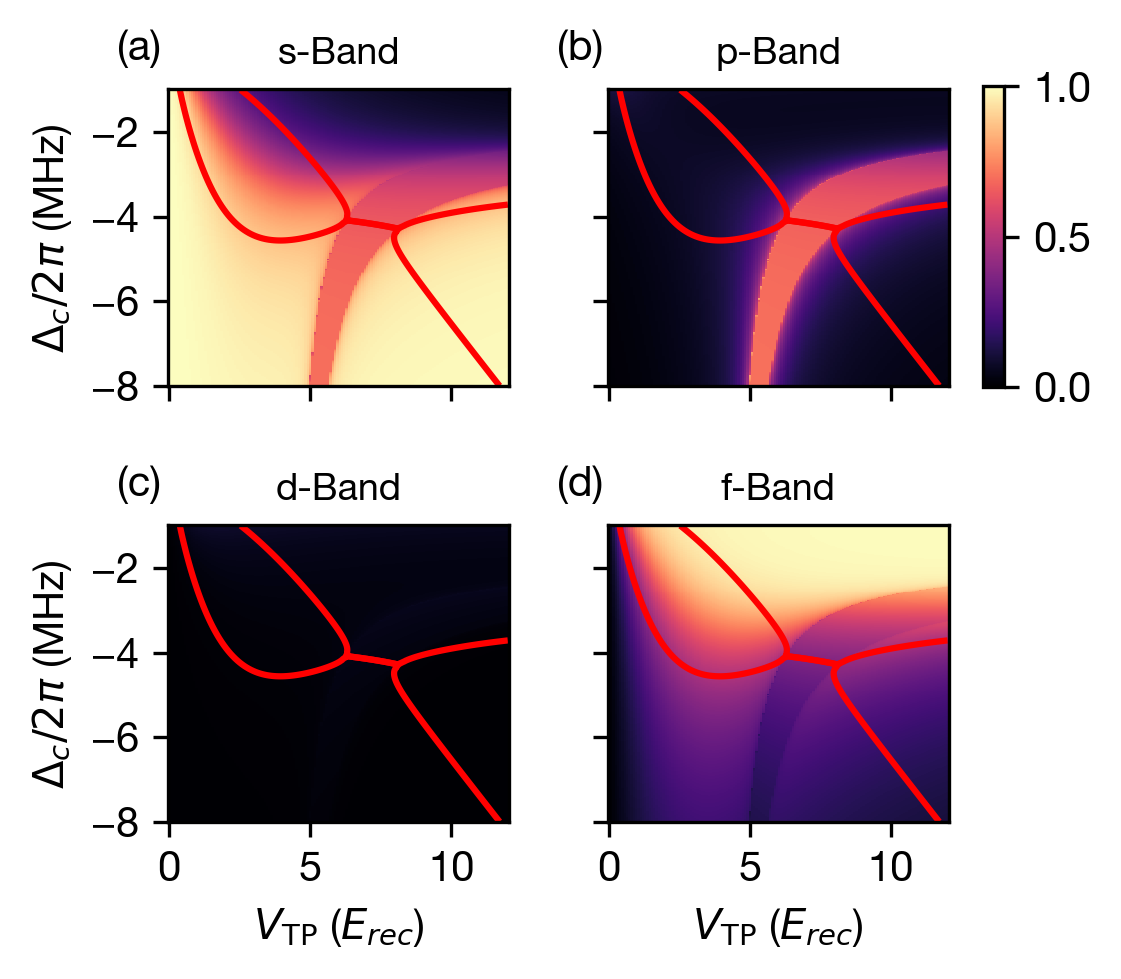}
\caption{\textbf{Decomposition of the precursor mode of SR1 in the open system setting.} The excited modes of the system projected back onto the Bloch-state basis, i.e. the four Bloch bands $(s,p,d,f)$ are displayed in panels (a)-(d). The crossing of an excited mode with zero energy is indicated by the red lines. The decomposition is qualitatively similar to the previous analysis, with a striking difference in the coalescence region. Here, the SR1 mode and the SR2 mode mix, which is evident, as the nature of the SR2 mode is a pure $p$ band in the absence of dissipative coupling (visible in panel (b)). Although the two modes begin to admix before they enter the coalescence region, this effect is localized around the coalescence region.}
\label{fig:decomp1:dissipative}
\end{figure}

As in section \ref{sec:natureOfmode}, we project the excitation modes back onto the different Bloch wave functions. In Fig.~\ref{fig:decomp1:dissipative} we show the decomposition of the SR1 mode. Outside the coalescence region, this shows qualitatively the same phase transitions and decompositions as for the hermitian case. However, within the coalescence area, we find a contribution of the SR2 mode, signalled by an admixture of the $p$ band. This can be expected, since the two modes fully mix before approaching the coalescence area. We therefore get a superposition of the typical SR1 pattern, which is dominated by a $\cos(\mathbf{k_- x})$ modulation, but has some residual $\cos(\mathbf{k_+ x})$ modulation together with the typical SR2 modulation, which is a checkerboard pattern consisting of $\sin(\mathbf{k_- x})$ and $\sin(\mathbf{k_+ x})$. In the coalescence area we find an equal mixture of $\sin(\mathbf{k_- x})$ and $\cos(\mathbf{k_- x})$, leading to a running wave along $\pm\mathbf{k_-}$. The sign of this running wave is determined by the complex phase between the two modes. The fact that the system chooses a direction is characteristic for the $\mathcal{PT}$-symmetry broken phase occurring in the coalescence area~\cite{miri2019exceptional,Heiss2012Physics}.

The analysis of the SR2 mode is analogous and the result is shown in Appendix~\ref{ap:dissipativeDecomp}. The SR2 decomposition outside of the coalescence region agrees with the closed-system case, and within this area we find the admixture of SR1 and SR2. 

\subsection{Nature of the coalescence}
In this subsection we investigate the physics occurring during the coalescence in more detail. The two excitation modes correspond to two distinct density patterns which oscillate in time at a frequency given by the mode energy. The cavity dissipation leads to an effective coupling between these two modes, which causes them to coalesce in a certain region. This coupling term can be seen in Eq.~\eqref{eq:dynamicalEqSecondorder} and is proportional to $\kappa\,\Xi_{0,i}\Theta_{0,j}$, where $i,j$ correspond to a band of SR1 and SR2, respectively.

We can map this situation to the synchronization of two dissipatively coupled harmonic oscillators, when tuning one of the resonance frequencies~\cite{Lu2023Synchronization}. For a range of $\Delta\omega$, i.e. the mismatch in resonance frequency of the two oscillators, they are locked in phase and oscillate at a common frequency. The extent of this synchronization range is determined by the strength of the dissipative coupling. During the coalescence, the degeneracy of the modes is lifted by the imaginary part of the energy, which is positive for one and negative for the other. These imaginary terms correspond to gain and loss in population of the respective modes when evolved in time. At the two points where the real and imaginary parts of the mode energies coincide, a spectral degeneracy occurs, where the two eigenvectors of the Hamiltonian become linearly dependent. This degeneracy is referred to as an exceptional point (EP), marking the transition between the $\mathcal{PT}$-symmetric and $\mathcal{PT}$-symmetry broken phase~\cite{heiss2004exceptional,Berry2004Nonhermitian}. Our many-body system exhibits the analogous phenomenology, where the role of the two harmonic oscillators is taken by the two collective excitation modes which are coupled via the dissipation of the cavity.

In the coalescence region, the two excitation modes evolve in time with a fixed relative phase. The periodic oscillation of these two phase-delayed, spatially-shifted density-patterns can then be recast into a moving pattern in a given direction. This agrees with our previous analysis of the mode decomposition, where a running wave component was found, and can again be related to the experimental observation of atomic pumping in this system, leading to transport~\cite{dreon2022}.

\begin{figure}[ht]
\centering
\includegraphics[width=0.48\textwidth]{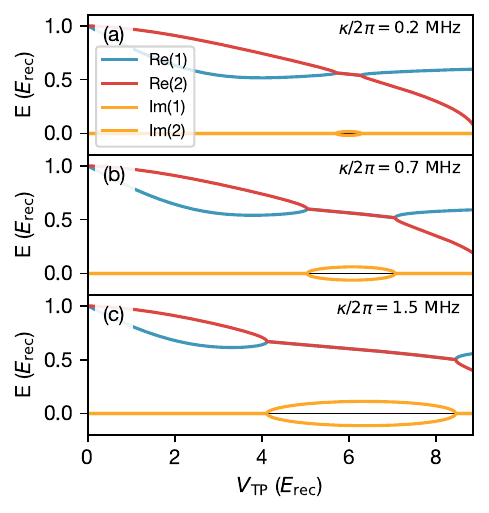}
\caption{\textbf{Mode coalescence for different dissipation rates $\kappa$.} Panels (a)-(c) show the energies of the SR1 and SR2 modes as function of $\Vtp$ for $\dc/2\pi=-5.2~\mathrm{MHz}$ and different values of cavity dissipation $\kappa/2\pi=0.2,0.7,1.5~\mathrm{MHz}$. The real parts of the two mode energies corresponding to SR1 and SR2 are shown in red and blue. The imaginary values are shown in orange and split into two components within the coalescence region. For increasing dissipation rates couplings, the coalescence region increases and larger gain and damping coefficients are obtained.}
\label{fig:Coalescence}
\end{figure}

In Fig.~\ref{fig:Coalescence}, we investigate mode coalescence in our system for different dissipation rates $\kappa$. The observed behaviour aligns with expectations for two dissipatively coupled harmonic oscillators: the excitation modes attract each other before merging at an exceptional point (EP), where their real energy components remain degenerate over the coalescence range, while their imaginary parts split into positive and negative components. Increasing the dissipative coupling extends the synchronization range of the two modes. This provides a framework for describing dissipative effects in our system, leading to mode synchronization and the emergence of EPs. Our model allows us to tune the dissipation constant and explore the nature of synchronized modes.

\section{Conclusion}
Using band theory, we developed a theoretical framework to explain the mechanism leading to mode softening and the nature of the excited modes in a transversally driven cavity-BEC system. We constructed a mean-field based model that is simple enough to allow for a clear physical interpretation of analytical expressions while still capturing the decomposition and softening of the excitation modes quantitatively.

We confirmed previous results regarding the superradiant phase 1 (SR1), describing it as a predominantly 1D density modulation along $\kmi$ for certain parameter regions~\cite{zupancic2019}. Our analysis extends this understanding by quantifying the residual density modulation along $\kpl$ and identifying the underlying physical origin of this 1D dominance -- namely, the energy competition between the $\kpl$ and $\kmi$ states and the kinetic energy of additional density modulations. We linked this competition to the angle occurring in our system, which allows to compare our model to setups featuring different angles between transverse pump and cavity field.

Building on this framework, we explored the effects of dissipation in our system. When dissipation is negligible, our open-system description aligns with closed-system predictions. However, in regimes where dissipative coupling dominates, we uncover a striking phenomenon: the coalescence of the two excitation modes. This coalescence is accompanied by mode mixing, the emergence of exceptional points (EPs), and the appearance of both amplified and damped modes, leading to chiral dynamics and the breaking of $\mathcal{PT}$-symmetry. By studying the effect of different cavity loss rates $\kappa$, we find good agreement with the experimental results presented in our companion paper~\cite{Natale2026Synchronization}.

We further linked the chiral dynamics in this phase to the atom pumping mechanism reported in Ref.~\cite{dreon2022}. However, a fully self-consistent treatment of the cavity field is expected to refine our quantitative predictions, particularly within the ordered phase. Our general framework can be applied to driven BEC-cavity systems across different geometrical configurations and cavity linewidths, and offers deeper insights into the nature of superradiant phases and their excitations. 

\section*{Acknowledgements}
We thank Tom Schmit, Erich Mueller and Daniel Pardo for insightful comments and suggestions on this manuscript.  This research was financially supported by the Swiss National Science Foundation (SNSF) project numbers\,186312, 212168, 217124, 221538, 223274, and the Swiss State Secretariat for Education, Research and Innovation (SERI) under grant number\,MB22.00090. This project is funded within the QuantERA II Programme, which has received funding from the EU's Horizon 2020 research and innovation programme under Grant Agreement No. 101017733.

\section*{Data availability}
The data that support the findings of this article are
openly available \cite{ResearchCollection}.

\bibliography{references}

\newpage
\begin{widetext}
\begin{appendix}

\section{System parameters}
We provide a summary of the system parameters used for the calculations performed in the main text, unless specified differently. 
\begin{table}[h]
    \centering
    \begin{tabular}{|c|c|c|}
        \hline
        \textbf{Parameter} & \textbf{Value} & \textbf{Units} \\
        \hline
        Imbalance parameter $\gamma$ & 1.3 & - \\\hline
        Atom number $N$ & $10^5$ & - \\\hline
        Cavity linewidth $\kappa/2\pi$ (section II\&III)& 147 & kHz \\\hline
        Cavity linewidth $\kappa/2\pi$ (section IV)& 400 & kHz \\
        \hline
        Atom cavity coupling $U_0/2\pi$& 47 & Hz \\
        \hline
        Atomic species & Rb$^{87}$ & - \\
        \hline
    \end{tabular}
    \caption{\textbf{System parameters used for numerical calculations.}}
    \label{tab:parameter}
\end{table}

\section{Numerical results for Bloch wavefunctions}
\label{ap:overlapPlots}
We show the results of the numerical calculation of the mode overlaps and mode energies, which are used in the approach outlined in the main text. We start with the energy of the four Bloch states located at $q=\mathbf{k_\pm}$, with band index $\alpha\in\{s,p,d,f\}$. In Fig.~\ref{fig:append:energyScaling} we show the energy $\omega_i(\Vtp)$ of these four states as a function of the transverse pump strength $\Vtp$. The two states in the $s$ and $p$ band associated with $\kmi$ start at $\SI{1}{\Er}$, the two states associated with $\kpl$ start at $\SI{3}{\Er}$. For increasing $\Vtp$, the lower two bands flatten, resulting in a reduction of energy of these states. The upper two bands also flatten, but simultaneously the band gap linearly increases the energy of these two modes.

\begin{figure}[ht]
\centering
\includegraphics[width=0.48\textwidth]{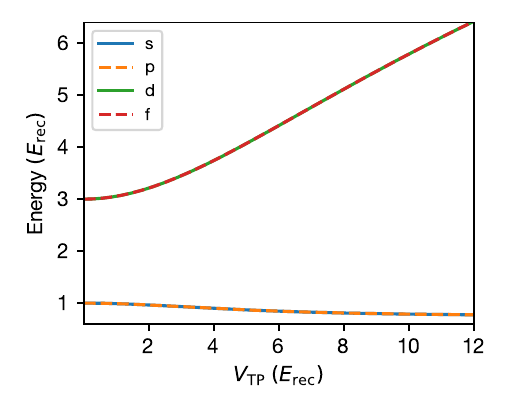}
\caption{\textbf{Energy scaling of the four Bloch states normalized to the BEC energy.} The energy of the four Bloch states located at $\kmi$ (or equivalently at $\kpl$) in the $s,p,d,f$ band as functions of the transverse pump power $\Vtp$.}
\label{fig:append:energyScaling}
\end{figure}

We calculate the overlap integrals $\Theta_{i,j}$ and $\Xi_{i,j}$ for the different Bloch functions $\phi_i(\mathbf{x})$,
\begin{align}
    \Theta_{i,j} &:= \frac{1}{A}\int_{W.S.}\phi^*_{i}(\mathbf{x})\cos(\kp\mathbf{x})\cos(\kc\mathbf{x})\phi_{j}(\mathbf{x})\mathbf{dx},\label{eq:app:overlapCC}\\
    \Xi_{i,j} &:= \frac{1}{A}\int_{W.S.}\phi^*_{i}(\mathbf{x})\sin(\kp\mathbf{x})\cos(\kc\mathbf{x})\phi_{j}(\mathbf{x})\mathbf{dx}\,.\label{eq:app:overlapSC}
\end{align}
As only overlaps between the BEC state and states at $\kpl$ and $\kmi$ result in a non-zero integral, we display only these overlaps: $\Theta_{0,i}$ and $\Xi_{0,i}$, where $i$ is one of the four bands $s,p,d,f$ at the BZ corner $\kmi$ (or equivalently $\kpl$).

\begin{figure}[ht]
\centering
\includegraphics[width=0.95\textwidth]{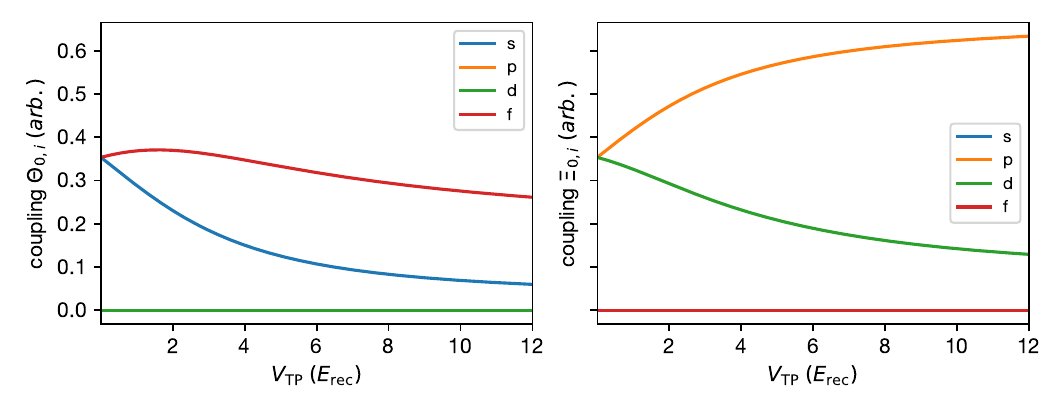}
\caption{\textbf{Overlap integrals  $\Theta_{0,i}$ and $\Xi_{0,i}$.} The overlaps $\Theta$ and $\Xi$ given in Eq.~\eqref{eq:app:overlapCC} and \eqref{eq:app:overlapSC} between the BEC wavefunction and different Bloch states at $q=\kmi$ in the $s,p,d,f$ band are displayed as a function of the transverse pump strength $\Vtp$. These overlap integrals deternine the coupling between the involved states.}
\label{fig:app:couplingsCCSC}
\end{figure}

As discussed in the main text and as can be seen in Fig.~\ref{fig:app:couplingsCCSC}, the $\coco$ potential only couples the $s$ and $f$ band, whereas the $\sico$ potential only couples the $p$ and $d$ band. For low $\Vtp$ the four non-zero coupling values start at the same value, as the transverse pump does not significantly modify the four Bloch states from simple $\sin$ and $\cos$ functions.

\begin{figure}[ht]
\centering
\includegraphics[width=0.65\textwidth]{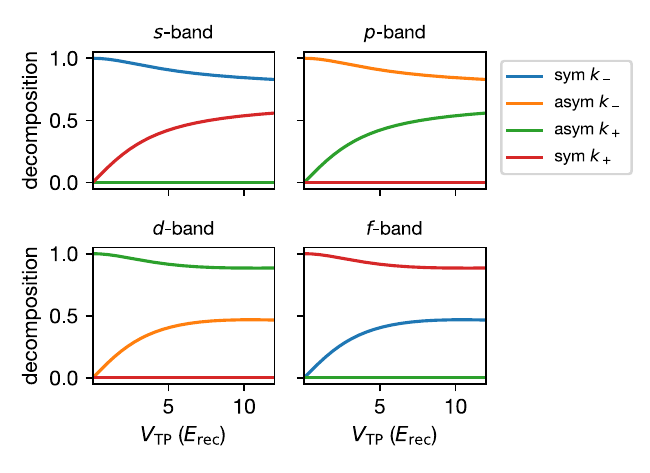}
\caption{\textbf{Decomposition of the four Bloch states into symmetry components.} The Bloch bands are projected on the symmetric (asymmetric) superposition of the four state: $\pm\kmi$ and $\pm\kpl$. Note, that as this is not a complete basis for high $\Vtp$, the different components are not normalized to 1 in this parameter range.}
\label{fig:app:blochDecomp}
\end{figure}

Finally, we discuss the decomposition of the four Bloch states into the symmetric and antisymmetric components in $\kmi$ and $\kpl$ direction. In general the four Bloch states can be written as superpositions of plane waves with different coefficients. For deep lattices, this expansion would contain high-momentum plane waves. To study the symmetry of these wavefunctions, we truncate the expansion to the lowest four momentum components: $\pm\kmi$ and $\pm\kpl$,
\begin{align}
    \phi_{s,\mathbf{k}_-}&\propto(\beta_0^s e^{i\mathbf{k_-x}}+\beta_1^s e^{-i\mathbf{k_-x}}- \beta_2^s e^{i\mathbf{k_+x}}- \beta_3^s e^{-i\mathbf{k_+x}})\\
    \phi_{p,\mathbf{k}_-}&\propto(\beta_0^p e^{i\mathbf{k_-x}}-\beta_1^p e^{-i\mathbf{k_-x}}-\beta_2^p e^{i\mathbf{k_+x}}+\beta_3^p e^{-i\mathbf{k_+x}})\\
    \phi_{d,\mathbf{k}_+}&\propto(\beta_0^d e^{i\mathbf{k_-x}}-\beta_1^d e^{-i\mathbf{k_-x}}+\beta_2^d e^{i\mathbf{k_+x}}-\beta_3^d e^{-i\mathbf{k_+x}})\\
    \phi_{e,\mathbf{k}_+}&\propto(\beta_0^e e^{i\mathbf{k_-x}}+\beta_1^e e^{-i\mathbf{k_-x}}+\beta_2^e e^{i\mathbf{k_+x}}+\beta_3^e e^{-i\mathbf{k_+x}}),
\end{align}
where $\beta_i^\alpha$ are arbitrary positive coefficients. We note, that the states in the $s$ and $f$ band share the symmetry with a cosine in both $\kmi$ and $\kpl$ direction, whereas for the $p$ and $d$ band this is the symmetry of a sine.  The decomposition of the Bloch states into symmetric and antisymmetric components in $\kmi$ and $\kpl$ direction can be seen in Fig.~\ref{fig:app:blochDecomp}.
For low values of $\Vtp$ the $s$ and $p$ band start out as pure density modulations in $\kmi$ direction, as this has a lower energy cost compared to a modulation in $\kpl$ direction. The $d$ and $f$ band instead start out as a $\kpl$ modulation, which has a higher energy cost. For higher values of $\Vtp$, the $\kpl$ modulation is admixed to the $s$ and $p$ band, resulting in a 2D checkerboard pattern. Note, that only the same symmetry components are admixed: for the $s$ band only symmetric superpositions of $\pm\kmi$ and $\pm\kpl$ are admixed. The four components do not add to $1$ for large $\Vtp$, as here higher momentum states start to get significantly occupied, which are not part of the basis shown in these plots.

\section{Bogoliubov spectrum calculation}
\label{ap:FreeEnergyFunctional}
As discussed in the main text, the mode energies can be found by a quadratic expansion of the effective Hamiltonian \ref{eq:Heffektiv} around the ground state outside the self-organized phase. Here, the system macroscopically occupies the BEC mode $\expval{\hat{c}_0}=\sqrt{N}$ and $\expval{\hat{c}_i}=0$, for $i\in\{1,..,4\}$. Using $\hat{c}_i=\expval{\hat{c}_i}+\delta \hat{c}_i$ and by moving to mean-field, we obtain 
\begin{align}
    &H_\mathrm{B} = \sum_{i\neq 0} \omega_i \delta c_i^* \delta c_i+\frac{\dc N}{2(\dc^2+\kappa^2)}\sum_{i,j\neq 0}\left(\delta_-^2\Xi_{i,0}\Xi_{j,0}+\delta_+^2\Theta_{i,0}\Theta_{j,0} \right)\left[\delta c_i^* \delta c_j+\delta c_i \delta c_j^*+\delta c_i^* \delta c_j^*+\delta c_i \delta c_j\right].
\end{align}

The explicit expression of the Bogoliubov matrix in the basis $\{\delta c_1,\delta c_2,\delta c_3,\delta c_4,\delta c_1^*,\delta c_2^*,\delta c_3^*,\delta c_4^*\}$ is 

\begin{align}
    \mathcal{L}=
\begin{pmatrix}
\mathrm{A} & \mathrm{B} \\
\mathrm{-B} & \mathrm{-A}
\end{pmatrix},
\end{align}
 with
\begin{align}
    &\mathrm{A_{i,j}}=\delta_{i,j}\omega_i+\frac{\dc N}{2(\dc^2+\kappa^2)}\left[\delta_-^2\Xi_{i,0}\Xi_{j,0}+\delta_+^2\Theta_{i,0}\Theta_{j,0}\right],\\
    &\mathrm{B_{i,j}}=\frac{\dc N}{2(\dc^2+\kappa^2)}\left[\delta_-^2\Xi_{i,0}\Xi_{j,0}+\delta_+^2\Theta_{i,0}\Theta_{j,0}\right].
\end{align}

\section{Mode decomposition with dissipation}
\label{ap:dissipativeDecomp}
For completeness we provide the decomposition of the SR2 mode for the case where cavity dissipation induces a coupling between the two modes. We see outside of the coalescence region a qualitatively similar mode decomposition as in the absence of dissipative coupling. In the coalescence region, the SR1 mode is admixed as visible from the non-zero occupation of the $s$ and $f$ band. 

\begin{figure}[ht]
\centering
\includegraphics[width=0.48\textwidth]{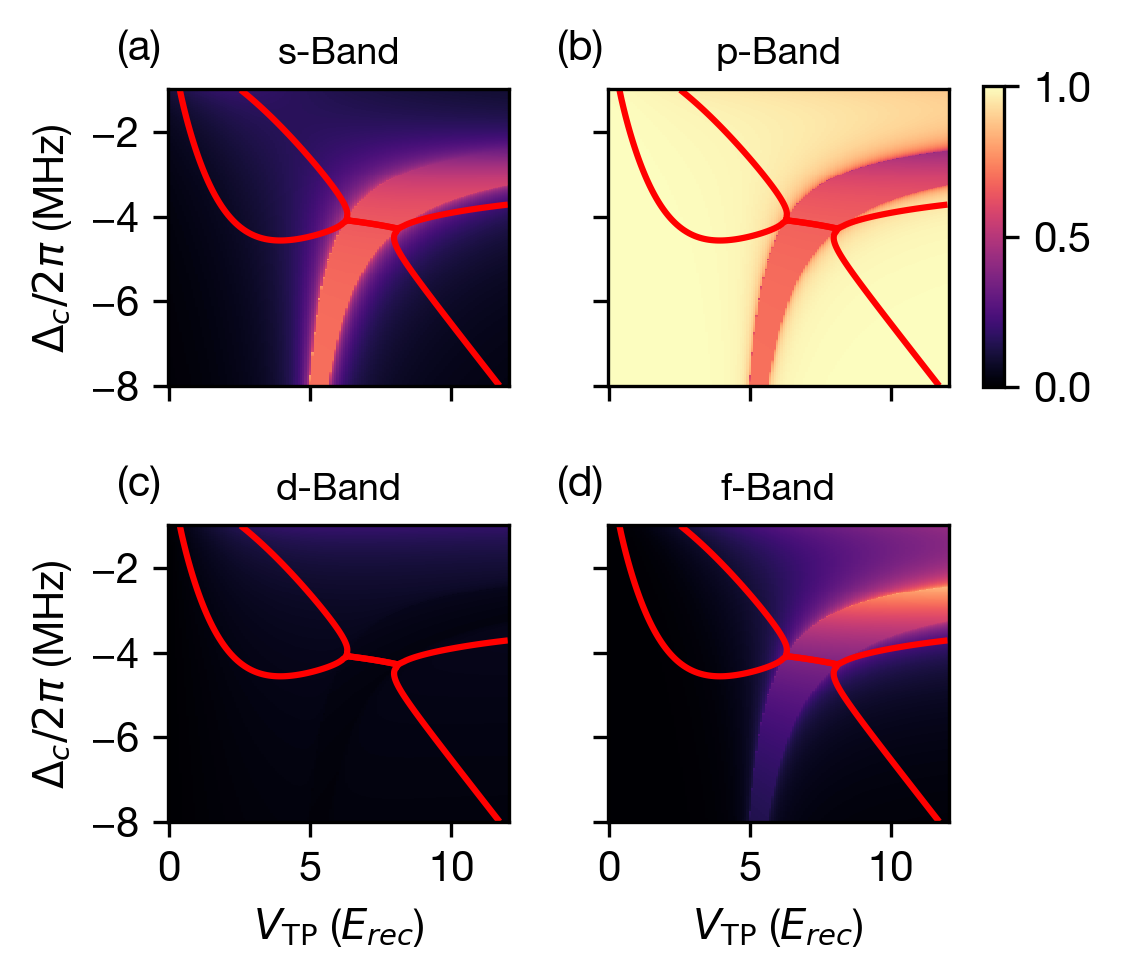}
\caption{\textbf{Decomposition of the SR2 mode in the open system setting.} As in the previous mean-field analysis, we project the excited modes of the system back onto the Bloch-state basis, describing the different symmetries. The red lines indicate where an excited mode crosses zero energy (i.e. the energy of the superfluid state). The decomposition is qualitatively similar to the previous analysis and as seen in  Fig.~\ref{fig:decomp1:dissipative} we get a different behaviour in the coalescence region. Here, the precursor mode of SR1 gets mixed to the mode of SR2.}
\label{fig:decomp2:dissipative}
\end{figure}

\end{appendix}
\end{widetext}

\end{document}